\documentclass[manuscript]{acmart}
\AtBeginDocument{%
  \providecommand\BibTeX{{%
    \normalfont B\kern-0.5em{\scshape i\kern-0.25em b}\kern-0.8em\TeX}}}

\acmDOI{}

\usepackage{algorithm2e} 
\usepackage{amsmath} 
\usepackage{array} 
\usepackage{balance} 
\usepackage{booktabs} 
\usepackage{caption} 
\usepackage{censor} 
\usepackage{colortbl} 
\usepackage{enumitem} 
\usepackage{fontawesome5} 
\usepackage{lipsum} 
\usepackage{listings} 
\usepackage{mdframed} 
\usepackage{multirow} 
\usepackage{nicematrix} 
\usepackage{pifont} 
\usepackage{rotating} 
\usepackage{setspace} 
\usepackage{siunitx} 
\usepackage{soul} 
\usepackage{stringstrings} 
\usepackage{subcaption} 
\usepackage{svg} 
\usepackage{svg-extract} 
\usepackage{tabularx} 
\usepackage{tcolorbox} 
\usepackage{tikz} 
\usepackage{url} 
\usepackage{verbatim} 
\usepackage{vcell} 
\usepackage{wasysym} 
\usepackage{wrapfig} 
\usepackage{xcolor} 
\usepackage{xspace} 

\newcolumntype{L}[1]{>{\raggedright\let\newline\\\arraybackslash}p{#1}} 
\newcolumntype{C}[1]{>{\centering\let\newline\\\arraybackslash}p{#1}} 
\newcolumntype{M}[1]{>{\centering\arraybackslash}m{#1}}

\definecolor{greenhighlight}{RGB}{194, 252, 186}
\definecolor{yellowhighlight}{RGB}{255, 255, 153}
\definecolor{bluehighlight}{RGB}{175,238,238}
\definecolor{orangehighlight}{RGB}{255, 188, 157}
\definecolor{alizarin}{rgb}{0.82, 0.1, 0.26}
\definecolor{MediumBlue}{HTML}{252dbc}
\definecolor{BlueGreen}{HTML}{117a65}
\definecolor{orangered}{HTML}{a93226}
\definecolor{aliceblue}{rgb}{0.94, 0.97, 1.0}
\definecolor{darkblue}{rgb}{0,0,0.7}

\newcommand{\citesec}[1]{Section~\ref{sec:#1}}

\newcommand{\etal}[1][]{\ifthenelse{\equal{#1}{}}{\emph{et~al.}\xspace}{\emph{et~al.}~\cite{#1}\xspace}}
\newcommand{\hf}{\textit{HIFuzz}\xspace}
\newcommand{\hfx}{\textit{HiFuzz's}}
\newcommand{\DFA}{\textit{D4A}\xspace}

\newcommand{\revchan}[1]{{\textcolor{black}{#1}}}

\newmdtheoremenv{definition}{Definition}

\lstdefinestyle{json}{
    basicstyle=\ttfamily\small\setstretch{0.8},
    showstringspaces=false,
    breaklines=true,
    frame=trlb,
    belowcaptionskip=1\baselineskip,
    xleftmargin=\parindent,
    morestring=[s]{"}{"},
    stringstyle=\color{darkblue},
    captionpos=b
}

\copyrightyear{2024}
\acmYear{2024}
\setcopyright{acmlicensed}\acmConference[CHI '24]{Proceedings of the CHI Conference on Human Factors in Computing Systems}{May 11--16, 2024}{Honolulu, HI, USA}
\acmBooktitle{Proceedings of the CHI Conference on Human Factors in Computing Systems (CHI '24), May 11--16, 2024, Honolulu, HI, USA}
\acmDOI{10.1145/3613904.3642958}
\acmISBN{979-8-4007-0330-0/24/05}
\begin{document}

\title{HIFuzz: Human Interaction Fuzzing for Small Unmanned Aerial Vehicles}

\author{Theodore Chambers}
\email{tchambe2@nd.edu}
\orcid{0000-0003-2923-3230}
\affiliation{%
  \institution{University of Notre Dame}
  \streetaddress{384A Fitzpatrick Hall}
  \city{South Bend}
  \country{USA}
  \postcode{46556}
}

\author{Michael Vierhauser}
\email{michael.vierhauser@jku.at}
\orcid{0000-0003-2672-9230}
\affiliation{%
  \institution{University of Innsbruck,  Department of Computer Science}
  \city{Innsbruck}
  \country{Austria}
}
\author{Ankit Agrawal}
\email{ankit.agrawal.1@slu.edu}
\orcid{1234-5678-9012}
\affiliation{%
  \institution{St. Louis University}
  \city{St. Louis}
  \state{Missouri}
  \country{USA}
}

\author{Michael Murphy}
\email{murphym18@gmail.com}
\affiliation{%
  \institution{University of Notre Dame}
  \streetaddress{South Bend}
  \city{Indiana}
  \country{USA}
}

\author{Jason Matthew Brauer}
\email{jbrauer90@gmail.com}
\affiliation{%
  \institution{Drone Response}
  \streetaddress{Denver}
  \city{Colorado}
  \country{USA}
  }

\author{Salil Purandare}
\email{salil@iastate.edu}
\author{Myra B. Cohen}
\email{mcohen@iastate.edu}
\affiliation{%
  \institution{Iowa State University}
  \city{Ames}
  \state{Iowa}
  \country{USA}
}

\author{Jane Cleland-Huang}
\email{JaneHuang@nd.edu}
\orcid{0000-0001-9436-5606}
\affiliation{%
  \institution{University of Notre Dame}
  \streetaddress{384A Fitzpatrick Hall}
  \city{South Bend}
  \country{USA}
  \postcode{46556}
}

\renewcommand{\shortauthors}{Chambers et al.}

\begin{abstract}
Small Unmanned Aerial Systems (sUAS) must meet rigorous safety standards when deployed in high-stress emergency response scenarios; however many reported accidents have involved humans in the loop. In this paper, we, therefore, present the HiFuzz testing framework, which uses fuzz testing to identify system vulnerabilities associated with human interactions. \revchan{HiFuzz includes three distinct levels that progress from a low-cost, limited-fidelity, large-scale, no-hazard environment, using fully simulated Proxy Human Agents, via an intermediate level, where proxy humans are replaced with real humans, to a high-stakes, high-cost, real-world environment. Through applying HiFuzz to an autonomous multi-sUAS system-under-test, we show that each test level serves a unique purpose in revealing vulnerabilities and making the system more robust with respect to human mistakes. While HiFuzz is designed for testing sUAS systems, we further discuss its potential for use in other Cyber-Physical Systems.}
\end{abstract}



\maketitle

\section{Introduction and Motivation}
\label{introduction}
\begin{figure}[t]
    \centering
       \begin{subfigure}[b]{.22\textwidth}
            \centering            
    \includegraphics[width=.90\textwidth]{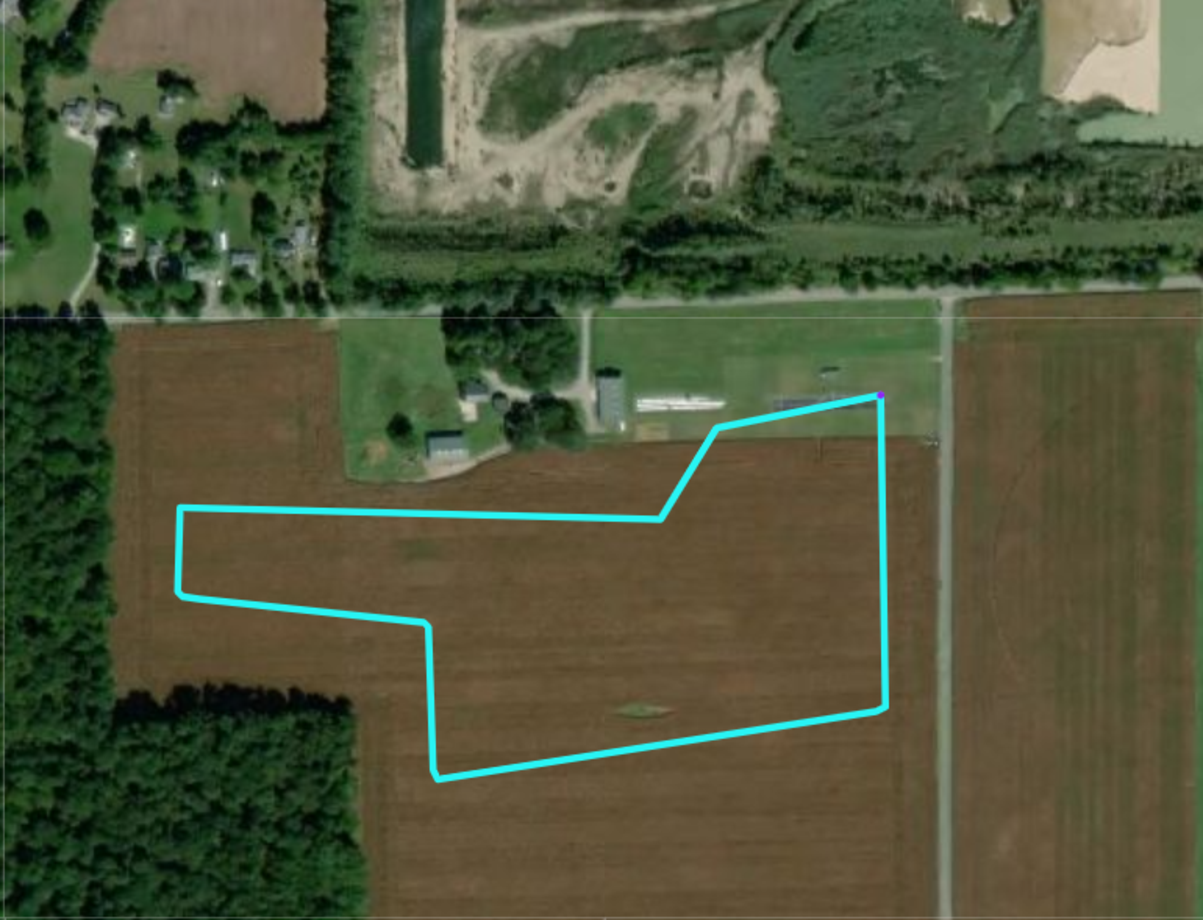} 
    \caption{The SW corner of the mission intersected with the geofence; however no geofence-action was set.}
    \end{subfigure}
    \hspace{9pt}
       \begin{subfigure}[b]{.22\textwidth}
            \centering            
    \includegraphics[width=.95\columnwidth]{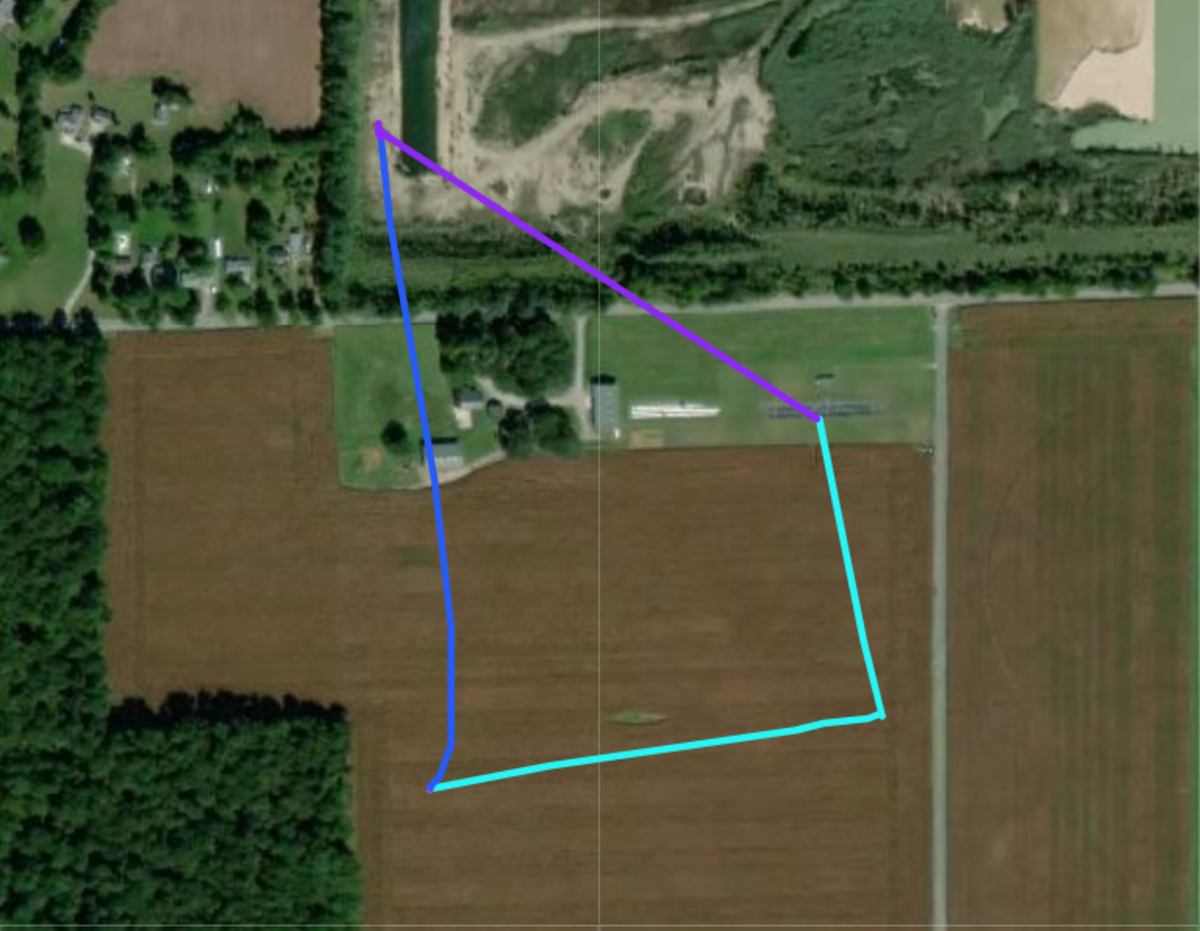}
    \caption{Flight replay showed that the sUAS flew north at increasing altitude after the geofence breach.}
    \label{fig:actualmission}
    \end{subfigure}
    \hspace{9pt}
    \begin{subfigure}[b]{.22\textwidth}
            \centering            
    \includegraphics[width=.95\columnwidth, clip, trim=0 10 0 0]{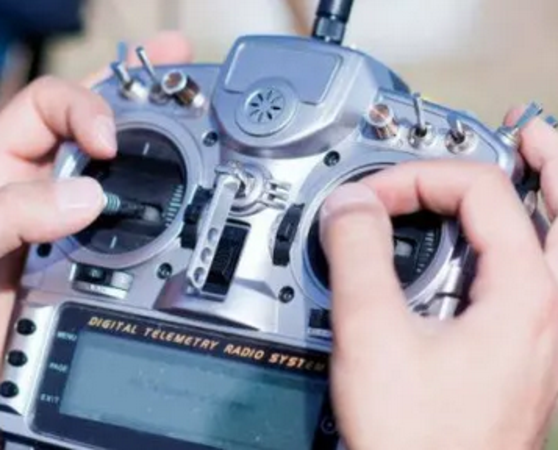}
    \caption{The RPIC must position the throttle correctly in case in-flight problems require human control. }
    \end{subfigure}
    \hspace{9pt}
    \begin{subfigure}[b]{.22\textwidth}
            \centering            
    \includegraphics[width=.95\columnwidth]{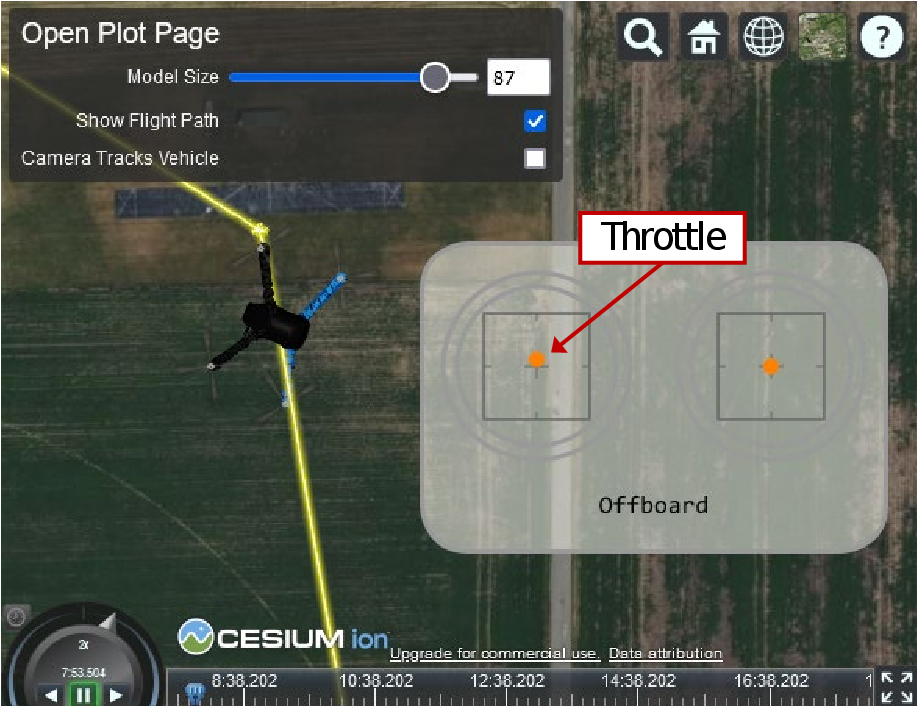} 
    \caption{Flight replay revealed that the RPIC had incorrectly set the throttle above neutral.\\ }
    \end{subfigure}
    \caption{Due to a combination of mistakes, including `operator error' by the Remote Pilot in Command, the sUAS flew off-route and ascended to 734 feet AGL. Note: All required regulatory reports were filed describing the incident.}
    \label{fig:oops}
    \Description{This set of figures shows snippets from an accident we experienced in Spring of 2023 with our own sUAS application. Figures a and b depict flight paths taken, and Figure c the radio transmitter used. Finally, Figure d shows the replay and analysis from the PX log analysis component.}
\end{figure}

Small Unmanned Aerial Systems (sUAS) need to meet rigorous safety requirements when deployed in high-stress emergency response scenarios~\cite{khan2019exploratory,herdel2022above}. However, the continual growth in sUAS deployment increases the risk of major incidents. Furthermore, several studies have reported that human ``errors'' have contributed to 65\% to 85\% of reported accidents in Cyber-Physical Systems (CPS) such as sUAS~\cite{nagel1998,kohn1999,fan2018simulation,vierhauser2021hazard}. We observed this phenomenon firsthand during a test flight in the Spring of 2023 (cf.~Figure~\ref{fig:oops}), when one of our autonomous sUAS breached a geofence, flew off its designated flight path, and ascended to an altitude of 734 feet above ground level (AGL) -- far above the legal limit of 400 feet AGL. A post-mortem analysis revealed a series of factors, including human-related missteps, that contributed to the incident. The remote pilot in charge (RPIC), who plays only a supervisory role under normal conditions, failed to set appropriate geofence-breach actions prior to the mission, placed the throttle in an incorrect position, lost situational awareness of the sUAS' trajectory following the geofence breach, and failed to take timely action when the sUAS started to fly off-course. However, blaming the operator for these accidents is very shortsighted. 

Human-Centered Design (HCD) focuses on creating and validating intuitive interfaces that are tailored to human cognitive capabilities~\cite{ucd1,ucd2} and, therefore, are designed to reduce human error. However, in the emergent area of sUAS, any failure to anticipate and address normal human ``mistakes''~\cite{endsley2012,endsley2017autonomous} can eventually lead to potentially dangerous incidents at critical moments of a flight. \revchan{A more systematic approach is therefore urgently needed to detect and mitigate design weaknesses that make the system vulnerable to human mistakes. In this paper, we propose human interaction testing techniques designed to reveal aspects of the system for which incorrect and unexpected human actions and inputs can result in potentially hazardous system behavior~
\cite{herdel2022above,luckeneder2017investigating,devries2018run}.}

We present the  {\bf H}uman-machine {\bf I}nteraction {\bf Fuzz} testing framework named ``\hf'', where ``HI'' represents both human interactions and the fact that sUAS fly at height. {\bf Fuzz} is analogous to traditional fuzz testing, where inputs are iteratively mutated and tested against the system to cover a large part of the behavior (and/or the code base) of an application~\cite{fuzz-survey:2022}, in order to reveal software defects and vulnerabilities~\cite{fuzz-survey:2022}. \revchan{Fuzz testing, also known as fuzzing, has been applied across various domains in software and system testing due to its effectiveness in uncovering vulnerabilities and defects~\cite{takanen2018fuzzing,fowler2018fuzz}; however, to the best of our knowledge, it has not previously been leveraged to probe for undesirable outcomes associated with human interactions.}

Our \hf framework includes three distinct levels (L1, L2, L3) progressing from a low-cost, limited-fidelity, large-scale, no-hazard environment, with fully simulated Proxy Human Agents (L1), via an intermediate level, where proxy humans are replaced with real humans (L2), to a high-stakes, high-cost, real-world environment (L3). \revchan{Replacing the human with a proxy in level L1 allows us to achieve fuzz-testing goals of rapid test coverage which would be impossible if a human were in the loop. At the same time, engaging humans in a small number of carefully selected tests at L2, allows us to investigate the human's situational awareness of the sUAS flight behavior ~\cite{endsley2012}. We can leverage this knowledge to identify appropriate design mitigations in the form of alerts, explanations, and even automated failsafe actions. Finally, level L3 further increases test fidelity by repeating tests that have successfully passed level L2, whilst introducing additional real-life stressors such as physical safety concerns and environmental detractors such as the glare of the sun, that are an inevitable part of field deployments.}

The levels are separated by two dedicated gateways. G1 resides between L1 and L2 and is responsible for down-selecting an appropriate set of tests to be executed in L2; while G2 represents a significant safety gateway in which standard safety assurance processes are followed, and hard decisions are made about executing \hf tests in the real world. Our tests are supported by tools for generating and executing the Fuzz Tests. For example, in the case of levels L2 and L3, where real humans participate in the tests, we have developed a mobile app to interactively guide users through the actions they need to perform during test execution. 

Our \hf framework makes three key research contributions. First, it presents a \emph{novel and systematic approach for human-interaction testing}, aimed at detecting, analyzing, and mitigating previously unknown hazards associated with human-sUAS interactions. Second, while \emph{Fuzz Testing} has been commonly used for software and systems tests, to the best of our knowledge it has not previously been used \emph{for human-interaction testing}. HiFuzz, therefore, makes a novel contribution, improving system robustness at the intersection of Human-Computer Interaction and Software Testing.  \revchan{Third, we conduct an in-depth analysis of \hf applied to our own multi-sUAS system, and a preliminary analysis of its generalizability across additional CPS. Results reported in this paper show that (1) \hf reveals system vulnerabilities associated with human interactions, potentially leading to their mitigation and improved design solution, (2) that all three test levels play a unique role in the testing process, and (3) that \hf can be applied across a broad range of CPS.} 

The remainder of the paper is structured as follows. Section \ref{sec:relwork} describes related work. Section~\ref{sec:singlefuzz}  explains how an individual fuzz test is specified, and Section ~\ref{sec:process} describes the various test levels and gateways. Sections \ref{sec:eval} and \ref{sec:analysis} describe experiments we conducted by applying \hf to a multi-sUAS system and provide a comprehensive discussion of the results. Finally, Section \ref{sec:threats} discusses limitations of our work, and Section \ref{sec:conclusions} draws conclusions.
\section{Related Work}
\label{sec:relwork}

In this section, we discuss related work associated with human-centered design of CPS, fuzz-testing in Software Engineering, human error and interaction in sUAS operations, and human interaction testing methodologies. Based on this prior work we argue that fuzz-testing can be an effective strategy for uncovering human-interaction vulnerabilities in the complex and dynamic CPS domains.

\subsection{Human Error and Interaction in sUAS Operations} Herdel~\etal[herdel2022above] conducted a comprehensive study focusing on over 100 applications across 16 diverse domains including emergency response and surveillance. They identified several research challenges pertaining to human-drone interactions, including one directly related to our work, addressing different ways in which people interact with sUAS to perform complex tasks. We address this issue through systematically testing outcomes of expected and unexpected human inputs for diverse tasks.

Rakotonarivo~\etal[rakotonarivo2023cleared] conducted interviews with drone operators, safety consultants, and regulators to identify operational risks and challenges when operating sUAS. One of their key recommendations was to ``Support exploration of operational parameters and estimate their impact on mission safety'' in order to allow ``operators to explore options that could simplify their procedures''. Our multi-level \hf process is designed to identify and mitigate potential safety issues before they arise in field testing, or worst-case, during live mission execution. It supports the systematic testing of diverse mission parameters and tasks and generates respective reports and documentation as inputs for subsequent safety analysis.

\revchan{Balot~\etal[balog2017examining] have collected a set of challenges associated with sUAS operations, related to HMIs, command and control, and management of sUAS operations. They argue that sUAS HMIs ``should be designed to take best advantage of human performance capabilities'', to ``[...] promote safety of flight operations''. While efforts have been taken to increase safety of sUAS operations~\cite{utm}, complex operational environments require thorough testing. This challenge was further investigated by Mccarley and Wickens~\cite{Mccarley_humanfactors} who proposed rules guiding levels of automation for different flight phases and operations and investigated different forms of control interfaces.
With \hf, we focus on this intersection in both simulated and real-world environments, by providing a thorough and structured multi-level testing framework.}


\subsection{Formal Methods for User Interaction Testing} 
\revchan{Several researchers have used formal methods to make mathematical claims about the correctness of the system with respect to user interactions, using a formal language such as temporal logic, a state machine, or process algebra~\cite{bolton2009method,curzon2007approach}. Diverse aspects of the system are modeled including expected outputs for given inputs, timing constraints, error handling requirements, the sequence of user interactions allowed by the UI, underlying state transitions, data flow and finally expected user behavior, including potential misuse or unexpected interactions \cite{formalui-1,formalui-2}. Formal verification techniques, such as model checking or theorem proving are then used to mathematically prove that the UI model satisfies the formal specifications, and meets the initially stated requirements and intended use cases. Formal models can also be used to generate test cases. For example, Bolton~\etal[bolton2013using] conducted a review on formal approaches in human-automation interaction. They showed that formal methods help to uncover potential shortcomings in human automation
interfaces, and are useful for diagnosing human-related system failures.
However, formal methods are only as good as the assumptions made during the specification and modeling process. In particular the models of expected user interactions including misuse cases, in an emergent area, such as sUAS are unlikely to be complete or correct.  \hf takes a somewhat orthogonal approach to formal methods, in that it assumes that the system is flawed, and probes the system to unearth these flaws.}

\subsection{Fuzz Testing in Software Engineering}

\revchan{
In the more general area of systems engineering, fuzz testing has emerged as an effective approach for testing large search spaces exhibiting high degrees of uncertainty (e.g., environmental factors)~\cite{chen2020active,wijaya2020domain}.}
\revchan{The majority of fuzzing techniques are greybox (using code-guided metrics to diversify coverage of program paths in the code)~\cite{padhye:2019,10.1145/2976749.2978428,10.1145/3580596,10.1145/3133956.3134020,10.1145/3324884.3418930}; however, scenario-based approaches, as adopted by \hf, represent an alternative approach for specification-based fuzzing \cite{lawbreaker:2022,10.1145/3395363.3397376,10.1145/3387940.3392252}}. Fuzzing has been used effectively within the CPS domain. For example, Kim~\etal[kim2019rvfuzzer] developed RVFuzzer to detect input validation bugs in robotic vehicle control programs including sUAS applications. However, they focused on detecting low-level controller malfunctions by monitoring vehicle control states.
Similarly, Kim~\etal[kim2021pgfuzz]~ created PGFUZZ, a policy-based fuzzing framework for robotic vehicles, and focused on safety and functional policies with respect to user inputs,  configuration parameters, and physical sUAS states. While they explicitly included user inputs and commands, they did not provide a comprehensive multi-level testing framework supported by safety analysis as used in \hf. Finally, Han~\etal[han2022control]~proposed a grey-box-based fuzzing
framework for detecting incorrect configurations in sUAS flight controllers. Their LGDFuzzer combined fuzzing with a genetic algorithm to detect potentially incorrect configurations and to test them in simulation, but did not consider human-related actions or real-world physical testing.

\section{Defining an Individual \hf Test}
\label{sec:singlefuzz}
Each individual \hf test focuses upon a human-interaction task that is conducted within a specific context.  In this section, we therefore describe the elements and properties used to define an individual test. 
\subsection{\hf Test Setup: Actors, Props, and Environment}
\subsubsection*{\bf Roles:~} Each human enacted task is assigned to a specific role such as a {\it Remote Pilot in Command} (RPIC), {\it Observer}~(OBS), {\it Mission Commander}~(MC), or {\it Safety Officer}~(SO).  We define \( \mathbf{R} \) as the set of roles represented by \( \mathbf{R} = \{r_1, r_2, r_3,  \ldots, r_i\} \). Depending on the current test level, either a human, or proxy-human assumes the assigned role.

\subsubsection*{\bf Interaction Devices:~} Humans perform a task using an interface device such as the radio control transmitter (RC), a GUI supported by a keyboard, mouse, and/or joystick, or another type of haptic device~\cite{neto2010high,yu2014human,mallan2017comparative,chen2021pinpointfly}. We define \( \mathbf{UI} \) as the set of all available user interfaces, represented as \( \mathbf{UI} = \{ui_1, ui_2, ui_3, \ldots, ui_j\} \).

\subsubsection*{\bf Drones and their Configurations:~} Tests can specify a specific drone or set of drones. Note that we utilize the word ``drone'', to emphasize the actual actual vehicle and its onboard flight controller, versus the complete software system.  
Inconsistencies across drones can cause accidents when their behavior fails to meet the human's current mental model~\cite{endsley2017autonomous}. We therefore define $\mathbb{D}$ as the set of drones, represented as \( \mathbf{D} = \{d_1, d_2, d_3, \ldots, d_l\} \).  Further, each drone in \( \mathbf{D} \) can be configured by the user prior to flight -{}- for example, by setting a geofence around the drone or assigning it a unique RTL (return to launch) flight altitude. We define $\mathbf{P}$ as a set of configurable parameters for an sUAS given by \( \mathbf{P} = \{p_1, p_2, p_3, \ldots, p_m\} \); however, low-level parameter configuration, that normally occurs when tuning the flight-controller~\cite{han2022control} is out of scope of this paper, and we assume that each drone has been adequately tuned and is flight-worthy. Parameters of interest are therefore limited to those exposed to the operator through interfaces (e.g., GUI screens) and therefore accessible during pre-flight setup.

\subsubsection*{\bf Simulation Environment:~} Finally, for Level L1 and L2 tests, depending upon the simulation environment used, we can directly configure elements such as wind. We define $\mathbf{E}$ as a set of configurable environmental parameters given by \( \mathbf{E} = \{e_1, e_2, e_3, \ldots, e_n\} \). 

\subsection{\hf Scripts} \label{sec:model-preconditions} Humans (serving in a specific role) enact a human-interaction task (HIT) in the context of an sUAS mission. Further, they execute the HIT when the sUAS and/or mission is in a specific state. For example, the RPIC might be asked to perform the action of switching to \textsc{position} mode when the drone is \textsc{flying} in \textsc{offboard} mode. This leads to the following specifications.
 
\subsubsection*{Missions:~}A mission represents the flight plans and other tasks that one or more sUAS will execute to provide context for the test. We define  $\mathbb{MSN}$ as the set of available missions, represented as \( \mathbf{MSN} = \{m_1, m_2, m_3, \ldots, m_q\} \). 

\subsubsection*{Human Interaction Task (HIT):~}There are two types of HIT that a human will perform during a test. First, the human could provide input to an individual sUAS through a hardware device such as the RC -- for example, by increasing the throttle, holding down the kill switch, or switching between modes. Second, the human could send a command to one or more sUAS via a GUI -{}- for example, issuing a global RTL command.
We define  $\mathbb{HIT}$ as the ordered set of interaction tasks performed by a user, represented as \( \mathbf{HIT} = \{hit_1, hit_2, hit_3, \ldots, hit_r\} \). 

However, CPS behavior is impacted by the current state of the system. Therefore, each HIT has an associated set of preconditions that also need to be defined. These preconditions are based on \textsc{modes},  \textsc{flight life-cycle states}, and \textsc{configurations}. Modes are used by almost every flight controller to support common flight tasks such as \textsc{take-off} and \textsc{loiter}, and to provide various degrees of flight stability (e.g., \textsc{stabilized} and \textsc{position-hold})~\cite{ardumodes, px4modes}. We define $\mathbf{M}$ as the set of flight modes, given by $\mathbf{M}$ = $\{m_1, m_2, m_3, \ldots, m_s\}$, where each mode \( m_i \) in \( \mathbf{M} \) is reachable in the SuT. We also define \( \mathbf{S} \) as a set of flight life-cycle states such as taking-off, flying, and landing, given by \( \mathbf{S} = \{s_1, s_2, s_3, \ldots, s_t\} \).  A drone can only be in one mode and one state at any time. Finally, we define configurations as the value assigned to any underlying parameter defined earlier as $\mathbf{P}$. Each HIT includes a mode and life-cycle precondition, and can optionally define a set of configuration parameters that serve as preconditions. Further, the precondition state must be reachable in at least one of the defined missions in order for any subsequent \hf test to be valid.

\begin{wrapfigure}[25]{r}{0.51\textwidth}
\vspace{-40pt} 
  \begin{lstlisting}[style=json, caption={A single test defined for the Sequence of Human Interaction tasks assigned two roles. The mission and environmental variables are shared across the entire test.}, label=lst:json_example]
{
  "Mission": "BASIC-WAYPOINTS",    
  "Environment": {
    "Wind": {
      "SPEED": "20KTS", 
      "DIRECTION": "NORTH"
    }
  },
  "Roles": [
    {
      "Role": "RPIC",
      "HITS": [
        {
          "ID": "1",
          "Drones": ["GREEN"],
          "Task": "MOVE THROTTLE TO +1",
          "Mode": "OFFBOARD",
          "State": "TAKING-OFF"
        },
        {
          "ID": "2"
          "Drones": ["GREEN"],
          "Task": "SET MODE TO STABILIZED",
          "Mode": "OFFBOARD",
          "State": "FLYING"
        }
      ],
      "Interaction_Device": "RC TRANSMITTER"
    },
    {
      "Role": "MC",
      "HITS": [
        {
          "ID": "1"
          "Drones": ["GREEN"],
          "Task": "PRESS RTL BUTTON",
          "Mode": "STABILIZED",
          "State": "FLYING"
        }
      ],
      "Interaction_Device": "GUI"
    }
  ]
}
  \end{lstlisting}
\end{wrapfigure}

\subsection{Defining the \hf Test}
Based on these definitions,  we can now specify an individual \hf test in a way that is sufficiently formal for automating test execution, but also readable to humans who serve as participants in the testing process.  We utilize JSON to represent each test as shown in Listing \ref{lst:json_example}. The test definition includes the mission, environmental factors, roles, the locally sequenced HITS, and preconditions performed by each role using a specific interaction device and drone. The \hf fuzzing engine ultimately uses these specifications to generate diverse combinations of properties, and the \hf Test Runner uses it to deploy the test, monitor its progress,  and to generate test prompts that are sent to the mobile app.

\subsection{Test Outcome}
\label{sec:test-outcome}
Each fuzz test is ultimately executed within the \hf platform, and its outcome is evaluated across two different dimensions -- first to determine if the test was {\it valid} or {\it invalid}, and second to determine if valid tests {\it passed} or {\it failed}. An invalid test fails to execute the full sequence of HITS, typically because preconditions for one or more of the HITS are never met. The outcome of valid tests is assessed as {\it passed} of {\it failed} based on {\it mission completion} and {\it mission adherence} criteria. 

\section{\hf Process: Test Levels and Gateways}
\label{sec:process}

\begin{figure}
    \centering
    \includegraphics[width=.85\textwidth]{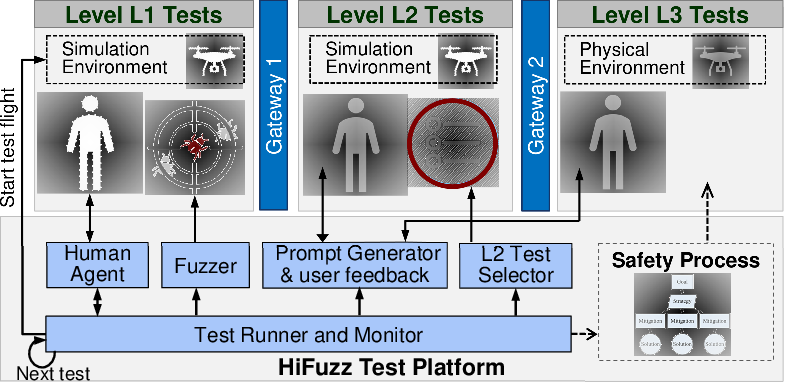}
    \caption{The \hf framework supports tests at all three levels. L1 operates fully in a simulated environment with support from a fuzzer and a proxy human agent. L2 operates with real humans in an otherwise simulated environment, and L3 operators in the physical world.}
    \label{fig:HiFuzzPlatform}
\end{figure}

The \hf process involves three testing stages (L1-L3) separated by two gateways (G1, G2), each of which serves a unique purpose (cf. Figure~\ref{fig:HiFuzzPlatform}). Individual tests are executed at each stage, however, the way they are executed, the role of human stakeholders, and the safety analysis that is performed prior to test execution differ greatly across stages.  In this section, we therefore describe each stage and gateway.

\subsection{Level L1: Large scale, simulated, fuzzing} The goal of L1 is to execute a large number of tests, as quickly as possible, without any of the risks involved in real-world sUAS flights. Therefore, L1  tests are run in the simulator using proxy human agents instead of humans. In the physical world, humans interact with sUAS via hardware devices, such as  RC transmitters, and their inputs are encoded into radio signals transmitted to the flight controller and transformed into flight commands (e.g., throttle, yaw, pitch, and roll adjustments, or mode changes). These inputs can be simulated through software-based, low-level function calls to the flight controller. Humans also interact with sUAS via GUIs, and these interactions can be simulated if the SuT exposes its API function calls. Utilizing these techniques, L1 is able to simulate human interactions (i.e., HITS) entirely in software, enabling thousands of fuzz tests to be run in a low-cost, low-effort, non-hazardous environment.

The L1 process starts with a planning task in which the \hf~ {\it tester} specifies the test features that constitute the fuzz space. As described in \ref{sec:singlefuzz}, these include roles, interaction devices, drones, environmental factors, missions, and HITs. The \hf fuzzer then uses this specification to automatically generate combinations of the defined properties and input values constrained by specific scenarios of interest. 
\revchan{The {\it Test Runner} iterates through the generated tests, invoking the mission in the simulation environment, monitoring the runtime state of each drone, checking for precondition states, and delegating HITs to the Proxy Human Agent when precondition states have been reached. The proxy mimics human input by replacing radio signals normally sent by the RC Transmitter, with MavROS manual control messages to simulate various switch changes and button presses for mode changes, throttle adjustments, and the kill switch.}  Results from each individual test are evaluated to determine if the test {\it passed}, {\it failed}, or was {\it untested} if the sUAS completed its mission without the preconditions ever being met.  All passed and failed outcomes are passed to Gateway G1.

L1 requires a simulation environment that accepts and executes a mission request -- potentially involving multiple drones, reports the progress of each drone throughout the mission, reports error messages, and produces a readable flight log at the end of each flight. 
Common examples of simulation environments that can be used to meet these requirements are Gazebo~\cite{gazebo}, jMAVSim~\cite{jMAVSim}, and AirSim~\cite{shah2018airsim}.



\subsection{Gateway G1: Downselecting for Human-in-the-Loop Tests}  G1 serves as a gateway between levels L1 and L2, and is responsible for selecting tests to be passed to L2. Its inputs are the tests and results from L1. It clusters these tests to identify groupings of similar inputs and outcomes, in order to guide the L2 test selection process. The number of clusters is based on budgeted L2 testing time or based on a standard approach such as the ``elbow-approach'' which looks for the sweet spot in terms of coupling and cohesion of clusters~\cite{syakur2018integration}. Typically, one or two representative tests are selected from each cluster for execution at level L2.

\subsection{Level L2: Humans in Simulated Environment} 
L2 tests are executed in the same simulation environment, however, humans replace the proxy agents, and interact with the sUAS through hardware devices (e.g., RC transmitters) and  GUIs used in physical deployments. As explained earlier, Level L2 is designed to provide higher degrees of fidelity than L1, while operating within a completely safe testing environment; however, it introduces higher testing costs with respect to human time and effort.
By integrating humans into the testing environment, L2 allows us to issue commands directly from the RC transmitter used in the field, providing increased fidelity of user inputs, and allowing direct observation of the sUAS behavior by human operators.  \revchan{Intuitively, Level L2 is needed to (1) execute a subset of interesting tests in a higher-fidelity environment, (2) to elicit feedback from humans about any failures that occurred in order to better understand their impact upon human operators, and ultimately (3) to evaluate the efficacy of user-facing mitigations, such as warnings or recommendations.}

\revchan{From a practical perspective, humans need help in determining when to perform a HIT, as many of the HIT's precondition states are internal, and not readily visible to human observers. \hf, therefore, provides a mobile app responsible for generating timely prompts. In order to minimize unnecessary mental overload of processing and responding to prompts, the Mobile App is designed with a simple GUI which gives the user planning time as well as clear instructions on what actions to perform. We designed and implemented the mobile app following principles of human-centered design, and our two test participants reported that it was intuitive and gave them clear and timely directions. However, a full assessment of the mobile app is outside the scope of this paper, and we therefore present it as a supporting tool rather than a primary contribution of this work.}

\begin{figure}[t]
    \centering
    \includegraphics[width=.9\textwidth]{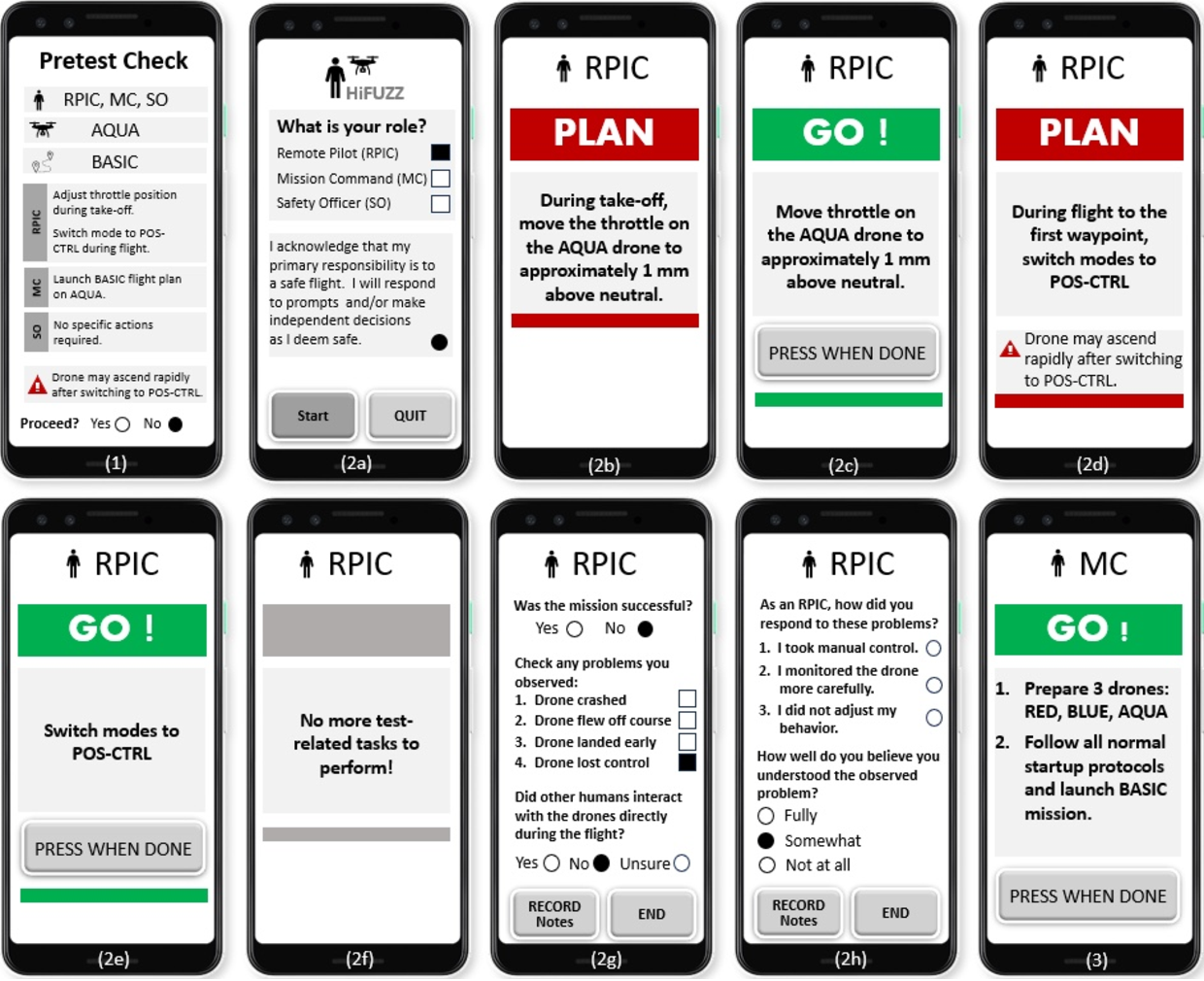}
    \caption{\hf Prompts are shared with human test participants via a mobile app. Here we show the design of the tester's precheck screen (1), followed by a series of prompts shared with the RPIC (2a-h), and MC (3) roles respectively. Figures represent the design which was fully implemented and deployed using React-Native.}
    \label{fig:mobile-app}
\end{figure}

A set of sample screens are depicted in Figure~\ref{fig:mobile-app}. The screens include preflight instructions and preparation (1, 2a, 3), a sequence of prompts that guide the RPIC (or other tester) through a sequence of tasks (2b-2f), and a series of post-test questions concerning the situational awareness of the operator (2g, 2h).  We only engage trained personnel in these tests, with the expectation (as required by regulations) that all participants are fully trained in their roles and know which switches and knobs to manipulate in order to execute the intended task. 

\subsection{Gateway G2: Safety Assessment and Mitigation}
While Test levels L1 and L2 seek to safely explore mission-breaking human-interaction faults that potentially cause erratic sUAS behavior, such as crashes and flight deviations, level L3's real-world deployment means that failures are potentially hazardous and costly. Therefore, Gateway G2 serves as a {\bf safety gateway} that ensures that each failed test from L2 is carefully assessed to determine if mitigations are needed, and that all tests deployed on the field with physical sUAS have undergone a rigorous hazard analysis with all identified hazards sufficiently mitigated. The aim is to (1) assess human-interaction vulnerabilities and flaws identified in levels L1 and L2, (2) mitigate them, (3) repeat level L2 tests to demonstrate that they have been successfully mitigated, and only then (4) proceed to level L3 tests. \hf does not dictate how the safety assessment should be performed as long as the process assesses hazards associated with each test case, e.g., using Fault-Tree Analysis (FTA) or Failure Mode Effect Criticality Analysis (FMEA/FMECA)~\cite{Reifer1979,sullivan1999galileo,leveson1983software}, evaluates mitigations to determine whether the risk has been satisfactorily addressed, and when needed, provides a semi-formal safety case, e.g., a Safety Assurance Case (SAC) that includes guidelines targeted at the human participants describing how the test can be safely conducted in the field.

\subsection{Level L3: Field Testing with humans-in-the-loop} \revchan{The goal at level L3 is to validate that all tests that have previously produced a failed L2 outcome have been demonstrably mitigated. Intuitively, real-world tests are essential for two reasons.  First, certain types of failures (especially race conditions) may only occur in the real world, and second, the human experience is different in the physical world than in simulation. For example, our own sUAS system was plagued for several months by a random take-off bug that appeared approximately once in every seven take-offs in the real world, but never in the simulator. Therefore, while simulations unearth many of the potential failures, repeated, real-world testing is essential for demonstrating that tests which executed successfully in simulation will also perform safely and correctly in the physical world.}

\subsection{\revchan{Assessing Test Outcomes}}
\revchan{\hf utilizes an ensemble of test oracles and techniques to determine whether each flight has been executed correctly.  These include analyzing runtime alerts generated by the flight controller and our own software system, reviewing mission logs, and considering human feedback received via the mobile app.  For the log analysis, we establish a ``blueprint'' representing an ideal mission outcome, and then use it as a point of comparison to measure deviation in the flight logs for each test. For each position timestamp in the blueprint we compute the distance to the nearest sUAS position in the current test log across the x, y, and z axes, and record the largest distance as the maximum observed deviation of the current log from the blueprint. We also extract other features from each log, such as the maximum altitude, the duration of the flight, the occurrence of free-falls, the final landing state, and the reported mission status throughout each mission.}
\begin{wrapfigure}[18]{r}{0.33\textwidth}
\vspace{1pt}
   {
    \begin{tcolorbox}[colback=gray!10, 
                      colframe=gray!20, 
                      boxrule=1pt, 
                      arc=4mm, 
                      left=6pt, 
                      right=6pt, 
                      top=6pt, 
                      bottom=6pt]
        \sffamily
        \small
        \setstretch{0.96}                      
        \textbf{Github Issue \#271:} Posted by: \censor{murphym18}, 08/03/2023 \vspace{3pt}\\     
        On August 3, during a flight test at \censor{peppermint field}, a near-crash incident occurred that exposed a safety issue in our current mission format. Right now we specify the altitude as meters above sea level. But due to several unlucky coincidences, on Aug 3rd the flight controller ended up with an incorrect altitude reading.\vspace{5pt}\\        
        When the drone was flying home, it ended up flying to a waypoint that was alarmingly close to the ground. This required immediate intervention by the Remote Pilot in Command (\censor{Jane}). The incident was not only intense and risky but could have resulted in significant damage.
    \end{tcolorbox}}
    \vspace{-12pt}
    \caption{An issue posted to Github describing a human-interaction incident, where the RPIC was forced to take control due to an altitude anomaly on the drone.}
    \label{fig:issue}
\end{wrapfigure}

\section{Experimentation: \hf Applied to \DFA System}
\label{sec:eval}
We evaluated \hf using a multi-sUAS system that we have developed and deployed in the real world as the {\it System-under-Test}. For double-blind review purposes, we refer to this as `Drones for All' (\DFA). Our evaluation focuses upon the outcomes of \hf rather than on the tools we have developed (i.e., the Mobile App), or the safety assessment (i.e., based on standard FMECA).  We address three research questions. 
\revchan{
\begin{itemize}
    \item[RQ1:] \emph{What kinds of human-interaction vulnerabilities were identified using the \hf process?}\\
    This question investigates the types of vulnerabilities detected using \hf.
    \item[RQ2:] \emph{Did each of the three test levels play a unique role in identifying human-related systems vulnerabilities?}\\
        This question explores the efficacy of the three test levels versus the additional costs of human-in-the-loop testing.
    \item[RQ3:] \emph{Is \hf generalizable across other human-intensive CPS applications?}\\
    This question takes a preliminary look at the generalizability of \hf to other domains.
\end{itemize}
}
The experiments described in this section were all executed in our \hf platform. 


\subsection{System under Test: \DFA}
\DFA is a distributed multi-user, multi-sUAS system, designed to support search-and-rescue~\cite{REMOVED}, aerial data collection, and surveillance activities~\cite{REMOVED}. Each sUAS is equipped with an \textit{Onboard Autonomous Pilot} (OAP) organized around a state machine which is dynamically configured for each mission. States support specific sUAS tasks such as {\it takeoff}, {\it search}, or {\it fly-to-waypoint} and vary greatly in complexity. For example, in the {\it takeoff state} the sUAS ascends to a predefined altitude and then transitions to a subsequent state such as {\it fly-to-waypoint}; while a {\it search} state utilizes AI-based computer vision capabilities to detect objects and make intelligent decisions, such as to {\it track} a person. A Ground Control Station (GCS) utilizes the MQTT message broker \cite{muccini2018iot} to coordinate system-level communication between sUAS, humans, and micro-services by publishing messages over a mesh radio. Status data (e.g., GPS location, battery, health) and task progress updates (e.g., current task, potential adaptations), are continually published by sUAS to support monitoring, analysis, and planning. Under normal operating conditions, humans set goals and send mission plans via GUI-based front-end clients; however, they can also directly issue commands via RC Transmitters. \footnote{If accepted we will provide a link to a video of the system in action.}

\subsection{Scenario-Based Fuzz Tests}
We adopted a scenario-based approach to test specific parts of the system. To select appropriate scenarios, we browsed through 272 issues (dated from 07/24/21 to 08/31/23) reported in the \DFA GitHub repository to identify incident reports associated with human-related incidents at the field (e.g., see Figure~\ref{fig:issue}). We selected two incidents as depicted in Table~\ref{tab:Scenarios}.

\definecolor{mediumgray}{gray}{0.6}
\definecolor{lightgray}{gray}{0.9}
\begin{table}[b]
 \caption{Two scenarios were selected in which human interactions were associated with flight failures. These scenarios were used in our experiments to drive scenario-based Fuzz Testing. The image for Scenario 1 is a replay from the physical flight logs capturing an actual incident that occurred on the flying field, while the image for Scenario 2 is taken from an L2 level simulation. Note: unless otherwise specified, modes and parameter values used below are described in the PX4 documentation \cite{px4modes,px4parameters}.}

    \label{tab:Scenarios}
    \centering
    \small
    \arrayrulecolor{lightgray}
    \begin{tabular}{|L{6.8cm}|L{.4cm}|L{6.8cm}|}

    \cline{1-1} \cline{3-3} 
    \cellcolor{black}\textcolor{white}{Scenario 1: Geofence breach} & &\cellcolor{black}\textcolor{white}{Scenario 2: Flawed take-over} \\
    \cline{1-1} \cline{3-3} 


    {\bf Observed:}~The drone hit the geofence with no geofence actions set and switched to stabilized mode. It then ascended rapidly and flew North. The RPIC had accidentally set the throttle just above neutral at the start of the flight. && 
    {\bf Observed:}~The RPIC took control of the drone by switching to STABILIZED mode with the throttle down. The drone oscillated as it attempted to stabilize and had a hard landing.\\
    \cline{1-1} \cline{3-3} 

    
    \includegraphics[width=6.8cm]{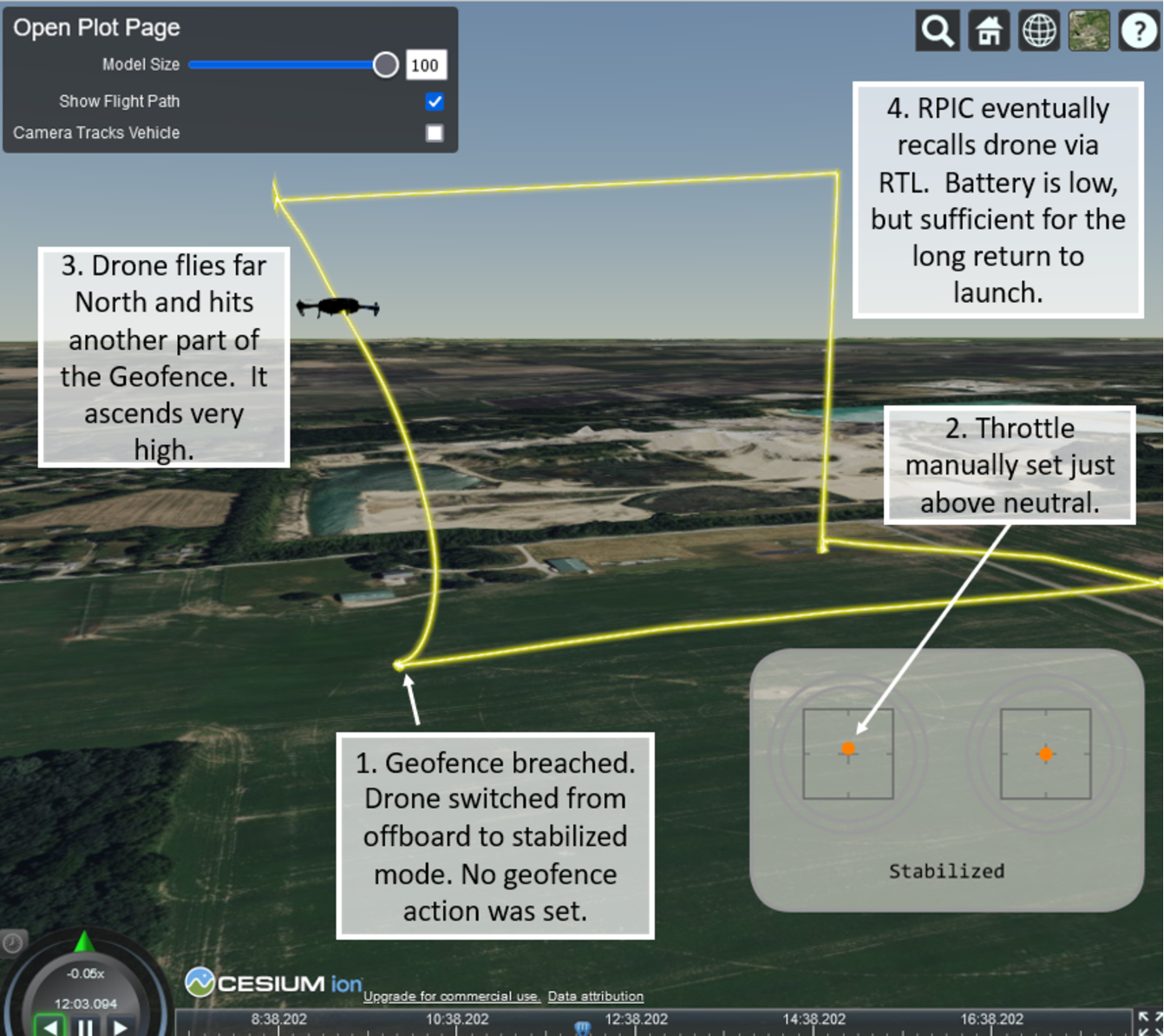} & &\includegraphics[width=6.8cm]{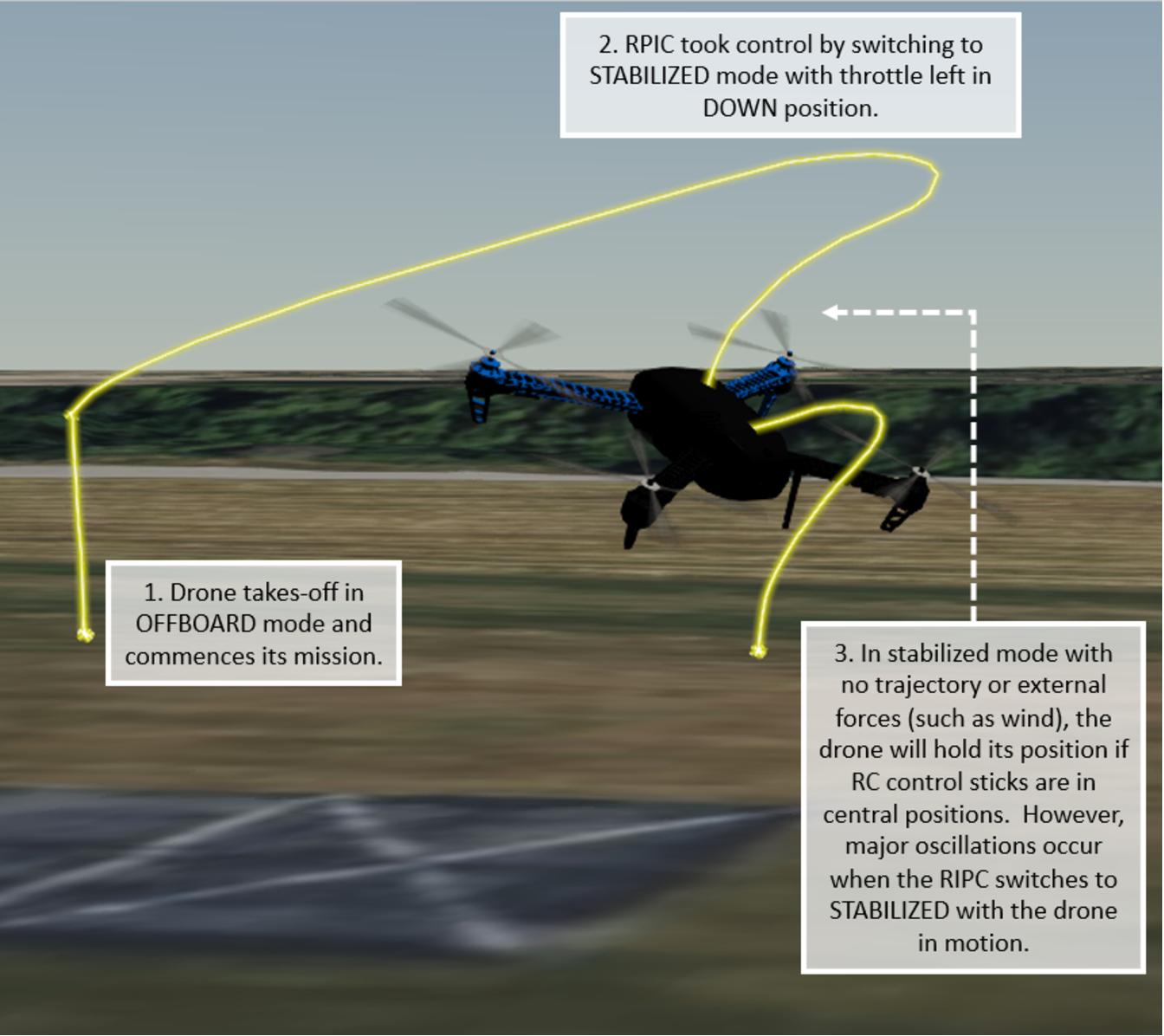} \\
    \cline{1-1} \cline{3-3} 

    
    {\bf Fuzz Scenario:}~Establish diverse geofences. Fly drone through geofence, where the RPIC defines geofence actions and sets throttle to various positions. As wind was suspected to be a factor, repeat in different wind conditions.&&
    
    {\bf Fuzz Scenario: }Test action of ceding control from the autonomous pilot to RPIC. Whilst drone is actively arming, disarming, or in the air, switch to POSITION mode with throttle in diverse positions. \\ 
    \cline{1-1} \cline{3-3} 




  \end{tabular}   
\end{table}

\subsection{Modeling the test space}
\revchan{We defined relevant properties as described in Section \ref{sec:singlefuzz}. For example, to test Scenario 1, we created a flight route that intersected a geofence. We defined the search space as all reachable modes and states, one drone (BLUE), one human role (RPIC), two types of wind, several properties associated with geofence settings, and several throttle settings. For all additional flight controller parameters, we accepted values defined during the drone's prior configuration process. Finally, we included three human actions (HITs) to (a) change mode, (b) adjust the throttle position, and (3) kill the motors (essential in case of dire emergencies or for failed takeoffs). This resulted in a test space of approximately 160,524 test configurations. We then systematically generated combinations of these properties and human actions (as explained in \ref{sec:singlefuzz}) and fuzzed the exact timing at which each action was to be executed once all test properties were satisfied. Finally, we created a simple flight test involving one drone taking off, flying to two waypoints, and returning home.}

\begin{table}[t]
    \centering
    \caption{Actual specification of the \hf fuzzing space used for experimentation purposes. Legend: blue= Initial states and modes are colored blue, yellow = configuration settings, orange=drones, green=human tasks. For Level L1 we only utilize the RPIC role and BLUE drone. Further Geofence\_Pred = `On' $\Rightarrow$ Geofence\_stat=`On' AND Geofence\_ACT $\Rightarrow$ Geofence\_Stat=`On'. This combination of features produced a test space of approximately 160,524 tests assuming no additional fuzzing around the precise timing of each test. }
    \label{tab:spec}
    \addtolength{\tabcolsep}{-3.2pt}
    \begin{tabular}{|c|c|c|c|c|c|c|c|c|c|c|c|c|}
     \cline{1-1} \cline{3-3} \cline{5-5} \cline{7-7} \cline{9-9} \cline{11-11} \cline{13-13}
\cellcolor{bluehighlight}{\bf Modes}&&\cellcolor{bluehighlight}{\bf States}&&\cellcolor{bluehighlight}{\bf Throttle POS}&&\cellcolor{yellowhighlight}{\bf Wind}&&\cellcolor{yellowhighlight}{\bf Geofence Act.}&&\cellcolor{greenhighlight}{\bf Roles}&&\cellcolor{greenhighlight}{\bf Human Tasks}\\  \cline{1-1} \cline{3-3} \cline{5-5} \cline{7-7} \cline{9-9} \cline{11-11} \cline{13-13}
ALTCTRL&&Pre-arm&&Maximum HIGH&&Medium Northerly&&0: None&&RPIC&&CHANGE-MODE\\  \cline{1-1} \cline{3-3} \cline{5-5} \cline{7-7} \cline{9-9} \cline{11-11} \cline{13-13}
POSCTRL&&Arm&&Medium HIGH&&High Northerly&&1: Warning&&\textcolor{gray!50}{MC}&&MOVE-THROTTLE\\  \cline{1-1} \cline{3-3} \cline{5-5} \cline{7-7} \cline{9-9} \cline{11-11} \cline{11-11} \cline{13-13}
OFFBOARD&&Takeoff&&Just above neutral&\multicolumn{2}{c}{}&&2: Hold mode&&\textcolor{gray!50}{SO}&&KILL-MOTORS\\  \cline{1-1} \cline{3-3} \cline{5-5} \cline{7-7} \cline{9-9} \cline{11-11} \cline{13-13}
STABILIZED&&Fly&&Neutral&&{\cellcolor{yellowhighlight}\bf Geofence Stat.}&&3: Return mode&\multicolumn{2}{c}{}\\  \cline{1-1} \cline{3-3} \cline{5-5} \cline{7-7} \cline{9-9} \cline{11-11}
AUTO.LOITER&&Hover&&Just below neutral&&On/Off&&4: Terminate&&{\cellcolor{orangehighlight}\bf Drones}\\  \cline{1-1} \cline{3-3} \cline{5-5} \cline{7-7} \cline{9-9} \cline{11-11}
AUTO.RTL&&Land&&Medium LOW&\multicolumn{3}{c|}{}&5: Land mode&&BLUE\\  \cline{1-1} \cline{3-3} \cline{5-5} \cline{7-7} \cline{9-9} \cline{11-11}
AUTO.LAND&\multicolumn{2}{c}{}&&Maximum LOW&&{\cellcolor{yellowhighlight}\bf Geofence Pred.}&\multicolumn{3}{c|}{}&\textcolor{gray!50}{ORANGE}\\  \cline{1-1}  \cline{5-5} \cline{7-7} \cline{11-11}

\multicolumn{6}{c|}{} &On/Off
&\multicolumn{2}{c}{}&&\textcolor{gray!50}{PURPLE}
\\ \cline{7-7}\cline{11-11}
    \end{tabular}
   
\end{table}

\subsection{Applying \hf to \DFA}
We executed all levels (L1 - L3) and gateways (G1, G2) for the planned scenario-based fuzzing of the \DFA system with the following setup.

\subsubsection{L1 Tests:} We ran 700 L1 tests based on various combinations of properties from Table~\ref{tab:spec}. Each test result was flagged with outcomes including the maximum altitude reached, flight duration, landed state, and mission completion. Any test exhibiting excessive altitudes, duration, excessively fast landing, or failure to complete the mission with final disarm, was labeled as ``Abnormal''. 

\subsubsection{G1 Gateway: } All tests in the profile were clustered
using Within-Cluster Sum of Squares (WCSS), using the elbow method to determine the number of clusters to be generated \cite{krzanowski1988criterion}. This ultimately resulted in nine unique clusters which were used as a guide to search for interesting test cases to pass to L2. For clusters containing at least one \textit{abnormal} test outcome, we selected the \textit{abnormal} test case that was closest to the centroid. We then inspected the profiles of tests close to the boundaries of each cluster in order to identify interesting edge cases.  This task took approximately one hour and resulted in the selection of 29 tests to pass to L2.

\begin{figure}[h]
    \centering
       \begin{subfigure}[t]{.28\textwidth}
            \centering            
    \includegraphics[width=.95\columnwidth]{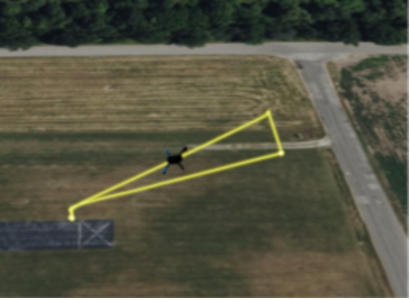} 
    \caption{The basic flight path of the sUAS when GEOFENCE=INACTIVE.}
    \end{subfigure}
    \hspace{4pt}
       \begin{subfigure}[t]{.28\textwidth}
            \centering            
    \includegraphics[width=.95\columnwidth]{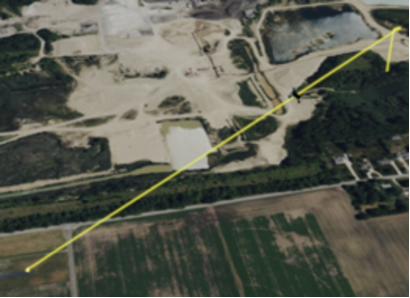}
    \caption{The actual flight path when Geofence=INACTIVE, and the RPIC executes a Mode Change to STABILIZED with DIR-Toggle="BACK".\newline}
    \label{fig:actualmissionfp}
    \end{subfigure}
    \hspace{4pt}
    \begin{subfigure}[t]{.38\textwidth}
            \centering            
    \includegraphics[width=.95\columnwidth]{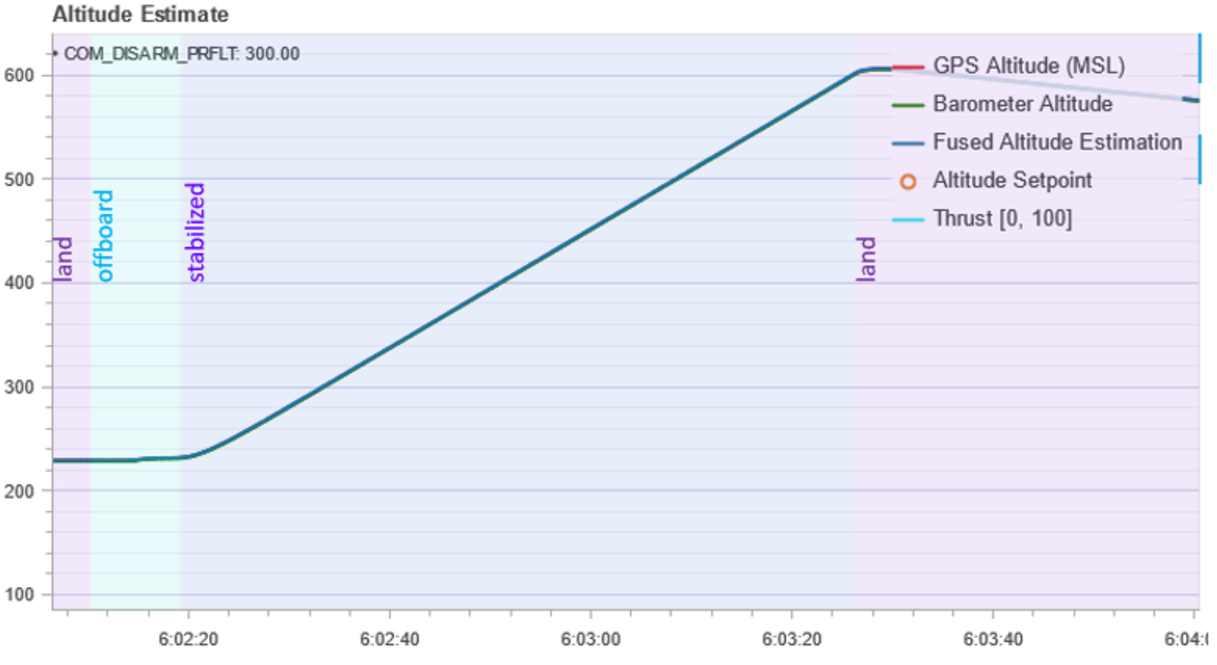} 
    \caption{By inspecting plots and log outputs we assess the outcome of the flight and identify root cause of errors. }
    \end{subfigure}
    \vspace{-12pt}
    \caption{In this case the RPIC switched modes to \textsc{stabilized} whilst the sUAS was flying in \textsc{offboard} mode. Due to the current trajectory and momentum of the sUAS, it continued its upward trajectory, ultimately reaching a height of 377 meters and a distance of over 550 meters. Ultimately, the TESTER issued a \textsc{land}  command to force an end to the mission. To minimize human errors caused by untimely mode-switches to \textsc{stabilized}, we can move the stabilized switch to a less prominent position, and add monitors to recognize if the drone is in 'free flight' due to a sudden switch to STABILIZE mode.}
    \label{fig:L2-analysis}
    \Description{This figure depicts examples from our tests with the resulting analysis performed in the px4 flight analysis platform. The shown analysis features provided information and visualization to identify the root cause of the problem for this case.}
\end{figure}

\subsubsection{L2 Tests: } Two researchers from our team executed all of the selected tests in the L2 simulation environment using a FrSky XD9 Plus Taranis Radio Handheld Controller~\cite{taranis}. The {\it tester} was responsible for the test setup, including launching the test runner, while the {\it RPIC} followed instructions displayed on the Mobile App, to conduct the planned human task at the correct stage of the mission. For each executed test, we preserved the flight logs, uploaded them into the PX4 flight log evaluation platform~\cite{px4logs}, then inspected the replayed flight log, logged messages, and graphs extracted from flight log data to further evaluate the flight outcomes. 
Figure~\ref{fig:L2-analysis} shows (a) the intended flight path of each test, (b) an actual flight path from one of the tests, and (c) one of the flight log data plots used to analyze the outcomes of a specific test. In this case, the  RPIC switched modes to \textsc{stabilized} (as directed by the test runner) whilst the sUAS was flying in \textsc{offboard} mode. Due to the current trajectory and momentum of the sUAS, the sUAS continued its upward trajectory, ultimately reaching a height of 377 meters and a distance of over 550 meters. The tester ultimately issued a \textsc{land} command to force an end to the mission.

\subsubsection{G2 Gateway: } \revchan{Two flight tests entered the G2 gateway during the course of our study. We  leveraged our existing safety analysis process to assess safety risks associated with executing them in the physical world, and constructed a safety case using the Goal Structuring Notation (GSN)~\cite{Kelly2004SAC}. Once the tests were deemed safe to deploy we placed them into the field-test backlog.  Due to space constraints, and the fact that the safety analysis process follows standard assurance practices, a deeper discussion on this gateway is out of scope of the paper.  When necessary, additional tests were written to validate specific mitigations.}

\subsubsection{L3 Tests:}
So far, we have only executed one L3 test in the field, which successfully validated that a previously revealed vulnerability from L1 and L2 had been successfully mitigated. We discuss this particular L3 test in Section \ref{sec:analysis}.  Other identified mitigations are currently backlogged in our development pipeline.

\section{Analysis of Results}
\label{sec:analysis}
We now discuss the results from our experiment with respect to each of the research questions. 

\subsection{RQ1: What kinds of human-interaction vulnerabilities were identified using the \hf process?
}

\revchan{To address this question we conducted a systematic inductive analysis of the L2 test results. As a first step, the four reviewers carefully analyzed each test case outcome, and marked the test as \textit{acceptable} or \textit{problematic}, where an acceptable test outcome was deemed to be one in which no problems were observed, and a problematic one included at least one undesirable outcome. All four reviewers agreed that nine cases were problematic and eight were acceptable; however, they held differing opinions on the remaining 12 and therefore engaged in discussions in order to reach consensus.  For example, there were three tests in which the RPIC pressed the kill switch to kill motors, but all three had different outcomes.  In one case, the sUAS landed immediately (desired behavior), in one case it performed an RTL (return to launch), and in a final case, it entered a tug-of-war with the sUAS' autonomous pilot and had a rather spectacular crash landing. Only the third test's outcome might be considered `bad', but in fact, the second case also was problematic as the observed behavior differed from expected.  It was therefore also labeled as problematic. These kinds of nuanced analyses are a known issue in Fuzz Testing -- where initial flags (passed/failed) tend to be rather coarsely applied.
Based on discussion between the four researchers, 10 tests were ultimately classified as acceptable (i.e., false positives selected at gateway G2), and 19 as problematic. }

\revchan{Each assessor also assigned a tag describing the problem from the human-interaction perspective. One researcher performed an initial card-sorting exercise on these tags to create named clusters, producing eight candidate groupings of human-interaction vulnerability types.  All four researchers then reviewed these groupings and discussed them in an online meeting. Following the discussion, six of the candidate groupings were retained (labeled 1-6 in Table \ref{tab:demons}), two groupings (\textit{fly-away} and \textit{failure to land}) were removed as they represented flight observations rather than human-interaction behaviors, and two additional categories were added (labeled 7-8 in Table \ref{tab:demons}). Table \ref{tab:demons} lists the number of failed tests by vulnerability types.}

\revchan{Some of the most common user interface design problems in CPS are related to poor Situational Awareness (SA), impacting the ability of users to perceive, understand, and to make effective decisions \cite{endsley2017autonomous}. These problems are documented as SA demons by Endsley \cite{endsley2012} with three additional ones identified by Agrawal et al.~\cite{Agrawal20}, as listed in Table \ref{tab:demons}. To gain deeper insights into the underlying design flaws we mapped each vulnerability to one or more relevant SA demon, and then leveraged these mappings as a useful resource for identifying meaningful mitigations.} 

\revchan{Here we describe one type of human-interaction vulnerability associated with \textit{incorrect stick positioning} (See Case \#1 from Table \ref{tab:demons}) as observed in five of the 29 test outcomes. Two of these cases involved incorrect throttle positions which is problematic if and when a human operator assumes manual control of the drone during flight. The problem originated from the default behavior of PX4 flight controllers, which requires the throttle to be fully down for arming. This behavior conflicts with the need for the throttle to be in the neutral position when the operator takes control so that the drone doesn't immediately crash land. We originally compensated for this problem by requiring the RPIC to move the throttle to the neutral position during takeoff in preparation for any later emergency. However, this created a stressful burden on the RPIC during a multi-sUAS takeoff. Our mappings to SA Demons associated the vulnerability with WAFOS (Workload, Anxiety, Fatigue, and Other Stressors) and MUI (transition failures across multiple interfaces) design demons. After gaining an understanding of the problem, we reprogrammed the takeoff routine to allow take-offs with the throttle in the neutral position thereby eliminating the previously required, error-prone human task. We also designed new alerts to warn the RPIC when the throttle was placed or left in a non-neutral position following takeoff. }

\revchan{Table \ref{tab:demons} depicts several other types of vulnerabilities that we identified through the inductive analysis. HIE-1 and HIE-2 represented cases in which failures repeatedly occurred due to expectations placed upon the human operators at high-pressure points in the timeline. Both were mitigated through automation thereby relieving humans from these high-stress, error-prone activities. HIE-3 and HIE-4 both revealed \textit{previously unknown vulnerabilities}. In HIE-3, the onboard autonomous pilot failed to recognize human interventions, thereby creating a tug-of-war between the human and the drone, leading to bizarre and unsafe flights; while in HIE-4, tests showed that the RC transmitter mappings included the ability for the operator to manually switch to offboard mode, meaning that the vehicle would no longer respond to commands from the RC transmitter. The remaining issues were all associated with loss of situational awareness related to a mode change. Brief descriptions are provided in Table \ref{tab:demons}.} 

\subsubsection{Types of Vulnerabilities}Based on this analysis we can answer RQ1.  The types of human-interaction vulnerabilities identified by \hf covered diverse areas of the system design. They included unrealistic expectations placed on operators to perform tasks under time pressure, affordances that allowed human operators to perform actions that they should not be able to do, and missing alerts that meant that operators often lost situational awareness. Furthermore, we found two cases (HIE-3 and HIE-4), which were entirely unanticipated vulnerabilities associated with human actions.  In the case of HIE-3, the tug-of-war detected by \hf was very similar to the root cause of Lion Air Flight 610 and Ethiopian Airlines Flight 302 in which the MCAS (Maneuvering Characteristics Augmentation System) incorrectly perceived the angle of attack to exceed predefined limits and therefore pushed the nose of the plane down, whilst pilots struggled to push it back up~\cite{boeing1,boeing2}. This demonstrates that the \hf process is capable of identifying highly critical and entirely unanticipated vulnerabilities.  Furthermore, in other cases, such as HIE-1, we had already observed related incidents in the field but had previously not fully understood the behavior. \hf tests provided new insights into the problem, leading to meaningful mitigations associated with automating prearming configurations and understanding when and where to issue warnings.

\subsection{RQ2: Did each of the three test levels play a unique role in identifying human-related systems vulnerabilities?}  

To answer this question we take a retrospective look at whether \hfx~three test levels all served a unique role. 
Level L1 tests were fully automated, not requiring human intervention, and answered questions such as ``did the flight complete successfully?'', and ``were there unexpected divergences from the planned route?''.  \revchan{However, we had to \textit{imagine} how an actual user would have observed and responded to the flight events that occurred. Therefore, even though \textbf{significant insights about potential human-interaction failures were gleaned from Level L1}, the results were insufficient for understanding users' perceptions and reactions to the problems as they occurred.  Drawing upon our previous example of the incorrect throttle position during takeoff, field tests showed that (1) the RPICs almost always adjusted the throttle, but (2) frequently placed the throttle in a slightly incorrect position, with large consequences. Feedback from RPICs clearly showed that these `mistakes' were due to stress and workload of supervising multiple sUAS during takeoff. A simple reminder would therefore be insufficient, and so we mitigated the problem through a complete redesign of the arming and takeoff routines, thereby removing this responsibility entirely from the operator. This type of insight is not obtainable with level L1 testing alone.} Further, while we have not yet conducted a full user study with the Mobile App we developed, in future work we will ask deeper questions of test participants concerning the current system and the efficacy of mitigations such as the use of specific alerts and recommendations. 

So far, this is one of only two tests that have been mitigated at L3. However, based on these two data points we observed that gateway G3 allowed us to take a deep dive into analyzing the safety concerns associated with executing tests in the field. It provided a safety net that helped us ensure that tests could be executed safely at Level L3. Demonstrating that the problem had been fixed and successfully deployed in the field built confidence that the system had satisfactorily addressed this particular system vulnerability. We conclude therefore that all three \hf levels provide critical support for human-interaction testing.

\begin{table}[]

  \addtolength{\tabcolsep}{-3.4pt}
    \centering
    \small

 \caption{\revchan{Mapping to Situational Awareness Demons}}  \label{tab:demons}
\begin{tabular}{|l|L{3.2cm}|C{0.4cm}|p{6.0cm}||
p{0.25cm}|p{0.25cm}|p{0.25cm}|p{0.25cm}|p{0.25cm}|p{0.25cm}
|p{0.25cm}|p{0.25cm}|p{0.25cm}|p{0.25cm}|p{0.25cm}|}
\hline
\multirow{4}{*}{HIE}&\multirow{4}{*}{Human Error Category}     & \multirow{4}{*}{\#} & \multirow{4}{*}{Outcome}   & \multicolumn{11}{c|}{SA Demon}\\\cline{5-15}
& & & &  
{\rotatebox[origin=c]{90}{AT}} & 
{\rotatebox[origin=c]{90}{MS}} & 
{\rotatebox[origin=c]{90}{IOL}} & 
{\rotatebox[origin=c]{90}{OLS}} & 
{\rotatebox[origin=c]{90}{EMM}} & 
{\rotatebox[origin=c]{90}{RMT}} & 
{\rotatebox[origin=c]{90}{WAFOS}} & 
{\rotatebox[origin=c]{90}{CC}} & 
{\rotatebox[origin=c]{90}{MUI}} & 
{\rotatebox[origin=c]{90}{STC}} & 
{\rotatebox[origin=c]{90}{EAU}} \\\hline


1& RC transmitter sticks set \newline incorrectly & 5 
&Unexpected flight behavior (e.g., ascends, descends, or flies off course after control is ceded to user).       
&    &   &    &  \multirow{2}{*}{\Circle}  &    &   &  \multirow{2}{*}{\CIRCLE}  &   & \multirow{2}{*}{\CIRCLE}&    &  \\\hline

2&Missing failsafe \newline configurations& 3 
& Operator fails to configure failsafes for each drone in the fleet in a consistent \& standard way.  
&    &   &   \multirow{2}{*}{\CIRCLE}  &    &  \multirow{2}{*}{\CIRCLE}   &   &   \multirow{2}{*}{\CIRCLE}  &   & &     & \\\hline
3&Human input ignored by \newline autonomous pilot&  7
& The autonomous system ignores a human-issued command, creating a ``tug-of-war''.                    
&    &   &    &  \multirow{2}{*}{\CIRCLE}   &    &   &    &  \multirow{2}{*}{\CIRCLE}  & &     &  \multirow{2}{*}{\CIRCLE} \\\hline

4& Inappropriate RC Switch \newline options  & 1 
& The RC transmitter switches are mapped to modes that the operator should not use.
&    &   &    &    &   &   &    &   & \multirow{2}{*}{\CIRCLE} &     &  \\\hline
5& Autonomous mode changes without notification& 3 & Human is unaware that the sUAS has switched mode and does not understand flight behavior.
&    &   &    &    & \multirow{2}{*}{\CIRCLE}   &   &    &   & &     &  \multirow{2}{*}{\CIRCLE}\\\hline

6& Inappropriately timed mode change by operator&4 
& Human changed to a mode that was inappropriate for current phase and state of the flight.
&    &   &    &    &  \multirow{2}{*}{\CIRCLE}   &   &    &  \multirow{2}{*}{\CIRCLE}  & &     &  \\\hline
7&Failure to operate drone according to its current mode & 4 
& Operator lacked or failed to apply appropriate piloting skills for current mode.
&    &   &    &    & \multirow{2}{*}{\CIRCLE}   &   &    &   & &     & \\\hline

8&Human loses situational awareness of sUAS behavior     &  6
&Complex series of events led to loss of situational awareness and inability to recover from a failure.
&    &   &    &  \multirow{2}{*}{\CIRCLE}  &    &   &  \multirow{2}{*}{\CIRCLE}  &   & & \multirow{2}{*}{\CIRCLE}    & \multirow{2}{*}{\CIRCLE} \\\hline

\end{tabular}
\vspace{3pt}
{\footnotesize

Legend: \emph{AT}=Attention tunneling, \emph{MS}=Misplaced Salience, \emph{IOL}=Information Overload, \emph{OLS}=Out of the loop syndrome, \emph{EMM}=Errant Mental Models, \emph{RMT}=Requisite Memory Trap, \emph{WAFOS}=Workload, Anxiety, Fatigue, \& other Stressors, \emph{CC}=Complexity Creep, \emph{MUI}=transition failures across Graphical \& Physical UIs, \emph{STC}=Socio-Technical CPS Communication Failure, \emph{EAU}=Enigmatic Autonomy. SG=Human Skill Gap.  \CIRCLE=Caused by, \Circle=Leads to.}
\end{table}

\subsection{RQ3: Is \hf generalizable across other human-intensive CPS applications?}

\revchan{While our \hf framework has been designed to identify risks related to human interactions in sUAS operations, its underlying concepts are applicable to a much broader range of CPS including other types of autonomous vehicles and ground-based robots.  \hf operates by fuzzing key system properties including (a) various modes in which a vehicle or robot operates, (b) different states it might transition into during the execution of a task or mission, and (c) potential human interactions with the system or robot. These core properties are found in other CPS, allowing \hf to be applied in other domains and for other types of system applications. To investigate the potential use of \hf across diverse CPS, we conducted a preliminary exercise of mapping the modes, states, and human interactions for systems from three different domains into \hf. These included a  centrally controlled sUAS system named Dronology, that used the Ardupilot Flight Controller \cite{dronology,cleland2018dronology}, a small robotic system developed by students to control a robot using a mobile phone, and a self-driving vehicle platform which we discuss in further detail.}

\revchan{The open-source, self-driving vehicle platform Autoware \cite{kato2015open,autoware26:online}  controls car operations and supports developers in creating autonomous car software systems. Similar to the modes available for our sUAS, Autoware manages different vehicle modes including Stop, Autonomous, Local, and Remote. Each of these modes represents a distinct operational setting for the vehicle. The \textit{Stop} mode halts all autonomous functions, while the \textit{Autonomous} mode enables full self-driving capabilities. \textit{Local} and \textit{Remote} modes refer to how humans interact with the car either with a steering wheel or over a network using a web application. An Autoware system can transition through multiple operational states such as \textit{Idle}, where the vehicle is not actively navigating; \textit{Active Navigation}, where the vehicle autonomously maneuvers through traffic or environments; and \textit{Emergency}, a state triggered during critical situations requiring immediate action or human intervention. Other states include \textit{Lane Following}, \textit{Lane Changing}, and \textit{Parking}. Further, the Autoware system also supports human intervention during vehicle operations, such as steering adjustments or mode switching.
Additionally, self-driving vehicles operate in different environmental conditions, such as rain, snow, and bad lighting, and hence require rigorous testing.
The concept of a \hf test (as defined in~\citesec{singlefuzz}) is therefore not unique to sUAS applications and potentially could be extended to other CPS that interact with humans and operate in a safety-critical, real-world environment.
While individual aspects of a system are domain-specific (e.g., a role might be the backup driver instead of an RPIC), its key elements (Roles, Interaction Devices, Tasks, Modes, etc.) are applicable across very diverse contexts. 
For example, CARLA~\cite{carla} provides a high-fidelity simulation environment for executing driving simulations with a multitude of configuration options.  Scenario-based tests, such as driving an autonomous car on the road, under controlled conditions, can provide the context for the \hf fuzzing.}

\revchan{Having defined properties for each of these three systems according to the types of properties used to define and execute \hf tests, we draw the preliminary conclusion that \hf is well suited to probing for human-interaction vulnerabilities across diverse CPS systems. Further, many parts of the \hf infrastructure are entirely reusable including the test-runner, the mobile app, and the G1 clustering analysis. However, other parts of the infrastructure will need to be customized to each application and/or domain.  These include adapters for interfacing with the simulation environments and metrics for evaluating acceptable versus problematic test outcomes. Primary adopters of \hf are therefore likely to be domain experts with the technical skills needed to test a complex safety-critical system.}

\section{Limitations and Future Work}
\label{sec:threats}
The research described in this paper is empirical in nature and is subject to three primary threats to validity. 

\revchan{First, our tests were limited to the RPIC, which is potentially the most challenging human role for operating sUAS; however, we need to extend the study to include other roles such as the MC (Mission Commander) and SO (Safety Officer), assign a more extensive set of human-interaction tasks, and study the perception of our stakeholders to identify further points of perceived vulnerabilities.  In addition, we plan to allow humans to interact more freely with the L2 simulation environment, and deal with a far broader set of emergency tasks including deviant flight behaviors. Their success at intervening could serve as an indicator of the robustness of the design with respect to human interactions.}

\revchan{Second, while we conducted a preliminary investigation into the generalizability of \hf, due to time constraints, we have not yet implemented \hf in these systems.  Instead, the experiments reported here focused on our own multi-sUAS system as the system-under-test.  In future work, we plan to run experiments in the application of \hf to other sUAS and CPS systems.}

\revchan{Third,  we claimed that human-in-the-loop tests are essential for understanding how humans perceive problems and potential mitigations. We built the mobile app to not only guide users through the testing process but also to collect data from them describing their experiences during the test. Future work is needed to conduct user studies with the mobile app to evaluate its effectiveness. } 

Finally, as previously mentioned, the L2 level, while fully functional, had less fidelity to the field than we had intended, primarily because libraries used to interface the radio signals with software-based PX4 simulations had some limitations. In future work, we plan to augment, or ultimately entirely replace the L2 layer with a Hardware-In-The-Loop layer in which a physical flight controller is integrated closely into the simulated environment. This would further increase test fidelity and allow the RC transmitter to communicate over radio signals directly with the PX4 controller. Overall, increasing fidelity would allow more robust human-interaction testing, and improve the overall fidelity of our \hf pipeline.

\section{Conclusions}
\label{sec:conclusions}
In this paper, we have presented the \hf testing framework for probing a system for human interaction vulnerabilities.  \revchan{The multi-level approach progresses from a low-cost, limited-fidelity, large-scale, no-hazard environment, with fully simulated Proxy Human Agents (L1), through an intermediate level, where proxy humans are replaced with real humans (L2), to a high-stakes, high-cost, real-world environment (L3). In this paper we have focused on the systematic application of each part of the \hf process, to identify human-interaction hazards so that we can design, implement, and validate mitigations.  The end goal is to increase the robustness of the system so that it is fault-tolerant to normal human errors}. 

\hf can be beneficial in two different ways.  \revchan{First, for testing individual systems, \hfx~ multi-level approach provides a safe pathway for detecting vulnerabilities associated with human interactions in the system under test. While deploying \hf for a new system is non-trivial, the return on investment in terms of human-interaction safety can make it worthwhile. Second, the lessons learned within a specific project can be documented and reused across other projects from similar domains, in order to help designers to avoid vulnerabilities in the first place. We therefore plan to extend the scope of our \hf tests, and document results in the form of a catalog.}  

In conclusion, results from applying \hf to our own system under test have shown it to be effective in identifying critical human-interaction vulnerabilities, thereby directly addressing the need for improved system safety and robustness.


\bibliographystyle{ACM-Reference-Format}
\bibliography{icse24,seams22}
\balance

\end{document}